\documentstyle[twoside,fleqn,espcrc2,epsf]{article}

\newcommand{\be}{\begin{equation}}
\newcommand{\ee}{\end{equation}}
\newcommand{\bea}{\begin{eqnarray}}
\newcommand{\eea}{\end{eqnarray}}
\newcommand{\pardis}{\langle \mu \rangle}

\title{Deconfining transition in Full QCD}

\author{
    J.M. Carmona\address[ZARA]{Departamento de F\'{\i}sica Te\'orica, 
    Universidad de Zaragoza, 50007 Zaragoza, Spain },
    M. D'Elia\address[GENO]{Dipartimento di Fisica dell'Universit{\`a} di
    Genova and INFN, I-16146, Genova, Italy},
    L. Del Debbio\address[PISA]{Dipartimento di Fisica dell'Universit\`a 
    and INFN, Via Buonarroti 2 Ed. C, I-56127 Pisa, Italy},
    A. Di Giacomo\addressmark[PISA] \thanks{Speaker at the 
    Conference (Email: digiaco@df.unipi.it). Partially supported by MIUR and 
    by EC, FMRX-CT97-0122}, 
    B. Lucini\address[OXFO]{Theoretical Physics, University of Oxford,
    1 Keble Road, OX1 3NP Oxford, UK},
    G. Paffuti\addressmark[PISA], C. Pica\addressmark[PISA]} 

\begin{document}

\begin{abstract}
We present evidence that in full QCD with two dynamical quarks
confinement is produced by dual superconductivity of the vacuum
as in the quenched theory. Preliminary information is obtained
on the nature of the deconfining transition.
\end{abstract}

\maketitle

\section{Introduction}
\label{sec:introduction}

A schematic phase diagram of full QCD with 2 dynamical flavours 
($m_u = m_d = m_q$) is shown in Fig.~1.
The upper part of the diagram is the deconfined phase, the lower part 
is confined. The line is determined by the maxima of a number of
susceptibilities~\cite{karsch,jlqcd}, including the susceptibility 
$\chi_L$ of the Polyakov line
\be
\chi_L = \int d^3 x \langle L(\vec{x},0) L^\dagger(\vec{0},0) \rangle
\ee
and the susceptibility $\chi_{\rm ch}$ of the chiral order parameter
\be
\chi_{\rm ch} = \int d^3 x \langle \bar{\psi}\psi(\vec{x},0) \bar{\psi}\psi (\vec{0},0) \rangle \; .
\ee
All of them have a maximum at the same value of $T$, for a given $m_q$,
which defines the line in Fig.~1.
For $m_q >$ 3 GeV, the maxima of $\chi_L$ diverge proportionally to 
the volume $V$,
indicating a first order transition. At $m_q = 0$ there are theoretical 
reasons and numerical indications that the transition is 
second order~\cite{piwi,karsch,jlqcd}.
At intermediate values of $m_q$ (tiny part of the line in  Fig.~1)
the susceptibilities do not diverge with $V$, and this is interpreted
as absence of a phase transition: the line would correspond to a crossover.

Across the transition the density of the free energy $F$ is a function
of the order parameters. The singularities of derivatives of 
$F$ are related to susceptibilities of the order parameters. In our case
$L$ is the order parameter only at $m_q = \infty$, $\bar{\psi} \psi$
only at $m_q = 0$. A good order parameter in the whole range
of $m_q$'s could lead to a different assignment for the order of the 
transition.

Such a parameter could be the disorder parameter $\pardis$ which describes
condensation of magnetic charges~\cite{ldd}. That parameter has been 
constructed 
and tested in quenched theory~\cite{I,II,III} and its definition can 
be extended to full QCD~\cite{IV}. In the spirit of $N_c \to \infty$ arguments
one expects that the mechanism of confinement is the same in quenched
and full QCD, quark loops being non leading in $1/N_c$ expansion. The 
operator $\mu$ creates a magnetic charge, as defined by some given
abelian projection. $\pardis \neq 0$ signals dual superconductivity.
In the quenched theory the specific choice of the abelian projection proves 
to be immaterial~\cite{III,adriano}. For the details about the definition
of $\mu$ we refer to~\cite{I,II,III,IV}.
For the quenched theory it is found that $\pardis \neq 0$ for 
$T < T_c$ and that $\pardis = 0$ for $T > T_c$, 
$\pardis \simeq \tau^\delta$ as ${T \to T_c^-}$, with 
$\tau = 1 - T/T_c$. 
A finite size scaling analysis of the infinite
volume limit yields $\delta = 0.20(3)$ for $SU(2)$; $\delta = 0.50(2)$
for $SU(3)$. The critical index $\nu$ has the values $\nu = 0.62(1)$
for $SU(2)$, in agreement with Ref.~\cite{karsch2}, 
$\nu = 0.33(1)$ for $SU(3)$, corresponding to a first order transition~\cite{fuku}.
We repeat the same analysis for 2 staggered flavours, using
the Wilson action for the pure gauge sector, and $12^3\times 4$,
$16^3\times 4$,$32^3\times 4$ lattices. The machines used are APEmille crates.
Part of the results are already published in Ref.~\cite{IV}.

\begin{figure}[t]
\label{diagram}
\vspace{-4pt}
\begin{center}
\leavevmode
\epsfxsize=76mm
\epsffile{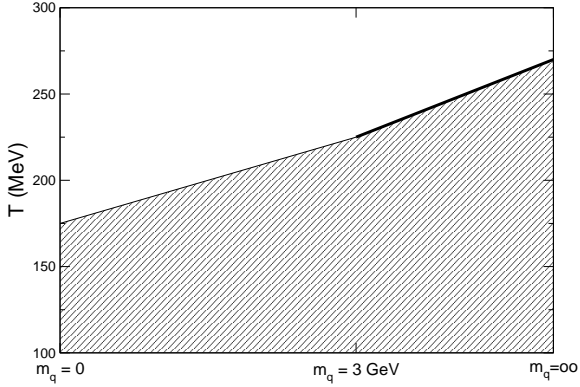}
\end{center}
\vspace{-35pt}
\caption{Phase diagram for two degenerate flavours, $m_u = m_d = m_q$.}
\vspace{-18pt}
\end{figure}

\begin{figure}[t]
\label{lowbeta}
\vspace{-4pt}
\begin{center}
\leavevmode
\epsfxsize=76mm
\epsffile{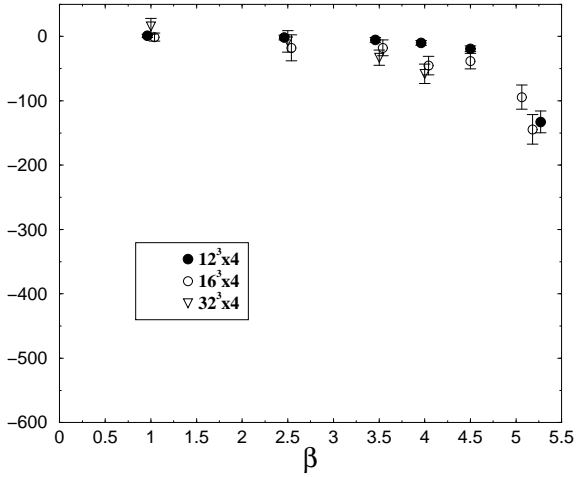}
\end{center}
\vspace{-35pt}
\caption{$\rho$ as a function of $N_s$ for $T < T_c$.}
\vspace{-18pt}
\end{figure}

\begin{figure}[t]
\label{highbeta}
\vspace{-4pt}
\begin{center}
\leavevmode
\epsfxsize=76mm
\epsffile{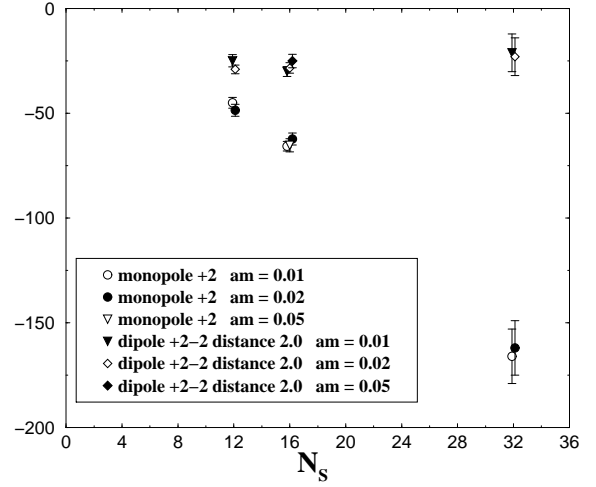}
\end{center}
\vspace{-35pt}
\caption{$\rho$ as a function of $N_s$ for $T > T_c$. As $N_s \to \infty$,
$\rho \to - \infty$ if the magnetic charge is non-zero and stays constant
otherwise.}
\vspace{-18pt}
\end{figure}

\section{Numerical Results}
\label{sec:numres}

As usual instead of $\pardis$ itself we determine the quantity
$\rho = \frac{\partial}{\partial \beta} \log \langle \mu \rangle$
in terms of which  
$\pardis = \exp \left( \int_0^\beta \rho(\beta') d \beta ' \right)$.
For $T < T_c$ we find that $\rho$ is practically size independent,
i.e. that $\pardis \neq 0$ as $N_s \to \infty$ (see Fig.~2).
For $T > T_c$ we find
(Fig.~3)
\be
\rho = -k N_s + {\rm const.}
\ee
i.e. that $\pardis$ is strictly zero as  $N_s \to \infty$. 
More extended checks of superselection of magnetic charge
in the deconfined phase have been done, showing that for different 
magnetic charges created by $\mu$,
$\rho \to -\infty$ if the net magnetic charge is non-zero, $\rho \to {\rm const.}$
if it is zero. Some examples are given in Fig.~3.

Around $T_c$ a finite size scaling analysis goes as follows. By dimensional
arguments one can parametrize $\pardis$ as 
\be
\pardis = 
\tau^\delta \phi\left( \frac{a}{\xi},\frac{N_s}{\xi},m_q N_s^\gamma \right)
\label{scaling}
\ee
where $a$ is the lattice spacing and $\tau = 1 - T/T_c$. If the correlation
length goes large, $\xi \sim \tau^{-\nu}$, $a/\xi \ll 1$ and
the dependence on $a$ can be neglected. The variable $N_s/\xi$ can be traded
for $\tau N_s^{1/\nu}$. In the quenched theory $m_q$ is absent and
$\pardis = \tau^\delta \bar{\phi}(0,\tau N_s^{1/\nu})$, which gives for $\rho$ 
the scaling behaviour $\rho/N_s^{1/\nu} = f(\tau N_s^{1/\nu})$, that 
can be tested and allows the determination of $T_c$, $\nu$ and 
$\delta$. In particular it implies that the height of the peak scales 
as $N_s^{1/\nu}$.

\begin{figure}[t]
\label{mpiconst}
\vspace{-4pt}
\begin{center}
\leavevmode
\epsfxsize=76mm
\epsffile{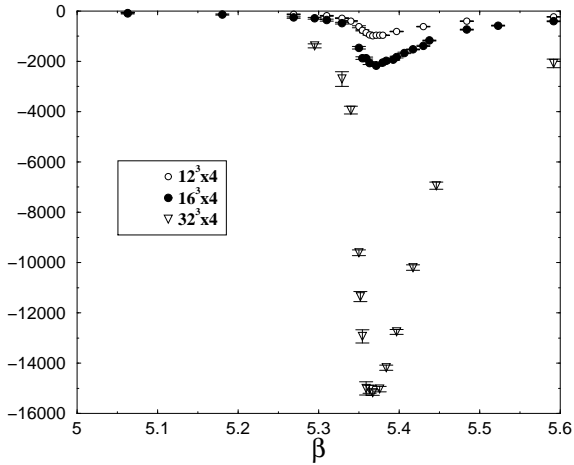}
\end{center}
\vspace{-35pt}
\caption{$\rho$ peak at different values of $N_s$ and $m_\pi/m_\rho \simeq 0.505$}
\vspace{-18pt}
\end{figure}

In presence of dynamical quarks, $m_q$ sets another scale, and we face a 
two scale problem. Simulations done at fixed ($m_\pi/m_\rho$)~\cite{IV}
show that the height of the peak $\rho$ roughly scales as $N_s^3 = V$
(see Fig.~4).
A more refined analysis can be done by choosing $m_q$ and $N_s$ 
such that $m_q N_s^\gamma$ in Eq.~(\ref{scaling}) is kept constant, 
and the problem is again reduced to a one scale problem, allowing 
to determine $\nu$, which gives information on the order of the transition.
The index $\gamma$ is known to be $\gamma \simeq 2.49$~\cite{karsch,jlqcd}, 
so that the three sets ($m_q = 0.075, N_s = 16$),
 ($m_q = 0.043, N_s = 20$) and ($m_q = 0.01335, N_s = 32$) 
keep $m_q N_s^\gamma$ constant. 
Preliminary results indicate
that $\nu = 1/3$, or that the transition is compatible with first order.
This is clearly visible from Fig.~5, where the height of the peak
is plotted versus $N_s^3$.
More detailed numerical analyses involving more values of $N_s$, $m_q$ and 
the use of an improved action, to improve the quality of scaling, are needed
to draw a definite conclusion.

\begin{figure}[h]
\label{rhopeak}
\vspace{-4pt}
\begin{center}
\leavevmode
\epsfxsize=79mm
\epsffile{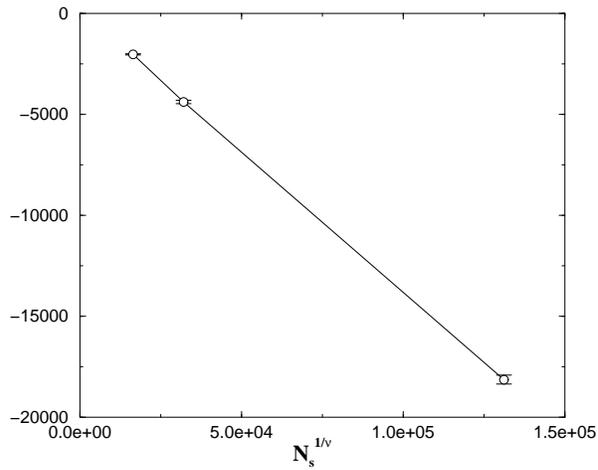}
\end{center}
\vspace{-35pt}
\caption{Values of $\rho$ at the peak as a function of $N_s^{1/\nu}$, with
$\nu = 1/3$. The height of the peak clearly scales as $N_s^{1/\nu}$.}
\vspace{-18pt}
\end{figure}

In any case it is demonstrated that also in full QCD
confinement is produced by monopole condensation.

\end{document}